\documentstyle[12pt,epsf]{article}
\expandafter\ifx\csname amssym12.def\endcsname\relax \else\endinput\fi
\expandafter\edef\csname amssym12.def\endcsname{%
       \catcode`\noexpand\@=\the\catcode`\@\space}
\catcode`\@=11
\def\undefine#1{\let#1\undefined}
\def\newsymbol#1#2#3#4#5{\let\next@\relax
 \ifnum#2=\@ne\let\next@\msafam@\else
 \ifnum#2=\tw@\let\next@\msbfam@\fi\fi
 \mathchardef#1="#3\next@#4#5}
\def\mathhexbox@#1#2#3{\relax
 \ifmmode\mathpalette{}{\m@th\mathchar"#1#2#3}%
 \else\leavevmode\hbox{$\m@th\mathchar"#1#2#3$}\fi}
\def\hexnumber@#1{\ifcase#1 0\or 1\or 2\or 3\or 4\or 5\or 6\or 7\or 8\or
 9\or A\or B\or C\or D\or E\or F\fi}
\font\tenmsa=msam10 scaled\magstep1
\font\sevenmsa=msam7 scaled\magstep1
\font\fivemsa=msam5 scaled\magstep1
\newfam\msafam
\textfont\msafam=\tenmsa
\scriptfont\msafam=\sevenmsa
\scriptscriptfont\msafam=\fivemsa
\edef\msafam@{\hexnumber@\msafam}
\mathchardef\dabar@"0\msafam@39
\def\dashrightarrow{\mathrel{\dabar@\dabar@\mathchar"0\msafam@4B}}
\def\dashleftarrow{\mathrel{\mathchar"0\msafam@4C\dabar@\dabar@}}

\def\ulcorner{\delimiter"4\msafam@70\msafam@70 }
\def\urcorner{\delimiter"5\msafam@71\msafam@71 }
\def\llcorner{\delimiter"4\msafam@78\msafam@78 }
\def\lrcorner{\delimiter"5\msafam@79\msafam@79 }
\def\yen{{\mathhexbox@\msafam@55 }}
\def\checkmark{{\mathhexbox@\msafam@58 }}
\def\circledR{{\mathhexbox@\msafam@72 }}
\def\maltese{{\mathhexbox@\msafam@7A }}
\font\tenmsb=msbm10 scaled\magstep1
\font\sevenmsb=msbm7 scaled\magstep1
\font\fivemsb=msbm5 scaled\magstep1
\newfam\msbfam
\textfont\msbfam=\tenmsb
\scriptfont\msbfam=\sevenmsb
\scriptscriptfont\msbfam=\fivemsb
\edef\msbfam@{\hexnumber@\msbfam}

\def\widehat#1{\setbox\z@\hbox{$\m@th#1$}%
 \ifdim\wd\z@>\tw@ em\mathaccent"0\msbfam@5B{#1}%
 \else\mathaccent"0362{#1}\fi}
\def\widetilde#1{\setbox\z@\hbox{$\m@th#1$}%
 \ifdim\wd\z@>\tw@ em\mathaccent"0\msbfam@5D{#1}%
 \else\mathaccent"0365{#1}\fi}
\font\teneufm=eufm10 scaled\magstep1
\font\seveneufm=eufm7 scaled\magstep1
\font\fiveeufm=eufm5 scaled\magstep1
\newfam\eufmfam
\textfont\eufmfam=\teneufm
\scriptfont\eufmfam=\seveneufm
\scriptscriptfont\eufmfam=\fiveeufm

\csname amssym.def\endcsname
\newif{\ifcomentarios}
\comentariosfalse
\renewcommand{\theequation}{\thesection.\arabic{equation}}

\newcommand{\be}{\begin{equation}}
\newcommand{\ee}{\end{equation}}
\newcommand{\bma}{\begin{displaymath}}
\newcommand{\ema}{\end{displaymath}}
\newcommand{\bc}{\begin{center}}
\newcommand{\ec}{\end{center}}
\newcommand{\text}{\rm}

\newcommand{\uflex}
{{\scriptstyle {\raise 9pt\hbox{$\backslash$}\,\!\!\!\!\!\Bigg\vert}}}
\newcommand{\ncm}{\newcommand}

\ncm{\rncm}{\renewcommand}
\ncm{\id}{{\bf 1}}
\ncm{\beq}{\begin{equation}}
\ncm{\eeq}{\end{equation}}
\ncm{\ba}{\begin{array}}
\ncm{\bea}{\begin{eqnarray}}
\ncm{\beanon}{\begin{eqnarray*}}
\ncm{\ea}{\end{array}}
\ncm{\eea}{\end{eqnarray}}
\ncm{\eeanon}{\end{eqnarray*}}
\ncm{\fns}{\footnotesize}
\ncm{\setc}[1]{\setcounter{equation}{#1}}
\newcounter{eqnr}
\newenvironment{eqnarrayabc}{\stepcounter{equation}
  \setcounter{eqnr}{\value{equation}}\setc{0}
  \rncm{\theequation}{\thesection.\arabic{eqnr}\alph{equation}}
  \begin{eqnarray}}{\end{eqnarray}\setc{\value{eqnr}}}
\ncm{\eqboxabc}[3]{\newline\parbox[t]{1.5cm}{#1}\hfill
  \parbox[b]{12cm}{\begin{eqnarray*} #3\end{eqnarray*}}\hfill
   \parbox[b]{1.5cm}{\vspace{-0.0cm}\begin{eqnarrayabc}#2\end{eqnarrayabc}}\newline}
\ncm{\eqbox}[2]{\newline\parbox{1.5cm}{#1}\hfill
  \parbox{12cm}{\beanon #2\eeanon}\hfill
  \parbox{1cm}{\bea\eea}\newline}
\ncm{\nr}[1]{\parbox{1cm}{\begin{eqnarrayabc}#1\end{eqnarrayabc}}\\}

\ncm{\kal}[1]{\mbox{$\cal #1 $}}
\ncm{\mrk}[1]{\!\!\! #1 \!\!\!} 
\ncm{\qed}{\hspace*{0.4cm}\rule{0.24cm}{0.24cm}}  
\ncm{\mbold}[1]{\mbox{\boldmath $ #1 $}}   
\ncm{\bm}{\mbold}
\ncm{\str}{\stackrel}
\ncm{\sub}{\subset}
\ncm{\e}{\varepsilon}
\ncm{\ka}{\kappa}
\ncm{\inputc}[1]{\begin{center}\input{#1}\end{center}}
\ncm{\lto}{\longrightarrow}
\ncm{\x}{\times}
\ncm{\bmm}{\bm{\cal M}}
\ncm{\cp}{{\bf P}}    
\ncm{\bfp}{{\bf P}}
\ncm{\bmi}{\bm{i}}
\ncm{\bmom}{\bm{\om}}
\ncm{\bmOm}{\bm{\Om}}
\ncm{\res}{\restriction}
\ncm{\bmL}{\bm{\cal L}}
\ncm{\bmell}{\bm{\ell}}
\ncm{\bmE}{\bm{\cal E}}
\ncm{\bme}{\bm{e}}
\ncm{\bmpi}{\bm{\pi}}
\ncm{\bmr}{\bm{r}}
\ncm{\bmsigma}{\bm{\sigma}}
\ncm{\wt}{\widetilde}
\newcommand{\beaa}{\begin{eqnarray}}
\newcommand{\eeaa}{\end{eqnarray}}

\begin{document}
\input{epsf.tex}

\author{{\bf Oscar Bolina}\thanks{Supported by FAPESP under grant
97/14430-2. {\bf E-mail:} bolina@lobata.math.ucdavis.edu} \\ 
Department of Mathematics\\
University of California, Davis\\
Davis, CA 95616-8633, USA\\
\and {\bf J. Rodrigo Parreira} \\ 
Instituto de Estudos Avan\c cados  \\
Rua Bar\~ao do Triunfo 375/304 \\
04602-000 S\~ao Paulo, SP \\
Brasil
}
\title{\vspace{-1in}
{\bf A Problem of Relative, Constrained Motion}}
\date{}
\maketitle
\begin{abstract}
\noindent
We develop a new method to determine the relative acceleration 
of a block sliding down along the face of a moving wedge. We
have been able to link the solution of this problem to that
of the inclined problem of elementary physics, thus providing 
a simpler solution to it.

\noindent
{\bf Key words:}  Newton's Laws, Constrained Systems,
Relative Motion, Friction, Relative Acceleration.\hfill \break
\end{abstract}

\noindent  
The problem of determining the relative motion of a block sliding down 
on the surface of a wedge which is itself free to move along a
frictionless horizontal plane \cite{{C},{L}}, as shown in {\it Fig. 1},
can be resolved by relating it to two problems of elementary Physics
having well-known solutions.
\newline
First suppose that the wedge is held fast and consider the problem of a
mass sliding down along the wedge's surface. This problem is equivalent
to the elementary problem of a mass sliding down a frictionless inclined
plane which makes an angle $\theta$ with the horizontal. The magnitude of
the acceleration $a_{m}$ of the mass moving down along the wedge's
surface is thus (\cite{APF}, p. 191)
\begin{equation}
a_{m}=g \sin\theta \label{am}
\end{equation}
Thus, when the acceleration of the wedge $a_{M}=0$, the block acquires
$a_{m}=g \sin\theta$ relative to the wedge.
\newline
A slightly more sophisticated variant of the situation described above 
occurs when one wants to find out what acceleration should be imparted 
to the wedge in order to keep the block from sliding down along the 
wedge's surface. In this case, the normal force of reaction {\it (N)}
of the wedge's surface on the block has a horizontal component which
makes the block move along with the wedge with acceleration $a_{M}$. 
Thus $N\sin{\theta}=ma_{M}$. On the other hand, since there is
no acceleration in the vertical direction, we have $N\cos{\theta}=mg$.
Eliminating {\it N} we obtain 
\begin{equation}
a_{M}=g \tan\theta \label{AM}
\end{equation}
Thus, the block stays at rest ($a_{m}=0$) relative to the wedge when
$a_{M}=g \tan\theta$.
\newline
The reader will note (\cite{APF}, p. 501) that when a simple pendulum is
suspended from the roof of a car moving with acceleration $a_{M}$ the
string hangs at an angle from the vertical which is given by (\ref{AM}). 
\newline
The two solutions ($a_{m}, a_{M})=(g \sin\theta,0 $) and ($a_{m},
a_{M})=(0, g\tan\theta$) provide an easy way to determine relationship
between the acceleration of the wedge and the acceleration of the block
relative to the wedge for any value of the acceleration imparted to the
wedge, from {\it zero} to $g\tan\theta$, where these extreme values
correspond to the solutions of the two limiting cases discussed above.
This is so because the variation of $a_{m}$ is directly proportional to
the normal force of reaction {\it N}, which, in turn, is also directly
proportional to $a_{M}$. Thus, the relationship between the accelerations
is a linear one, the pair of values ($a_{m}, a_{M}$) given above are two
points on a straight line as shown in {\it Fig. 2}. 
\newline
From this figure we get the general relationship between the accelerations
\begin{equation}
a_{m}=g \sin\theta - a_{M} \cos\theta \label{REL}.
\end{equation}
Our point is that Eq. (\ref{REL}) also holds when the wedge moves
solely under the weight of the sliding block, without any external
force imparting an acceleration to the wedge. In this case, the linear
momentum of the system (block and wedge) along the horizontal direction
is conserved, that is:
\begin{equation}
p=(M+m)v_{M} + m v_{m} \cos\theta \label{LM}
\end{equation}
where $v_{M}$ is the velocity of the wedge relative to the floor and
$v_{m}$ is the velocity of the block relative to the wedge.
\newline
Eq. (\ref{LM}) implies this second relationship between the
acceleration:
\begin{equation}
(M+m)a_{M} + m a_{m} \cos\theta=0, \label{ACE}
\end{equation}  
which is easily derived from the geometry of the system as in 
\cite{BV} and \cite{PARS}. 
\newline
From (\ref{REL}) and (\ref{ACE}) we solve the problem completely for
$a_{M}$ and $a_{m}$ \cite{C}:
\begin{equation}
a_{M}=-\frac{m g \sin\theta \cos\theta}{M + m \sin^{2} \theta}
\end{equation} \label{S1}
and 
\begin{equation}
a_{m}=\frac{(M+m) g \sin \theta}{M + m \sin^{2} \theta} \label{S2}
\end{equation}
\noindent 
\vskip .2 cm
\noindent
Even when there is friction between the block and the wedge an analogous
relationship between the accelerations can be easily obtained by the same
reasoning we have developed before, with minor changes to take friction
into account.
\newline
Let $\mu$ be the coefficient of friction between the block and
the wedge. Suppose also that $\mu < \tan\theta$.
\newline
When $a_{M}=0$ (the wedge is held fast again) the block acquires
an acceleration (\cite{H}, p.72)
\begin{equation}
a_{m}=g(\sin\theta - \mu \cos\theta)
\end{equation} 
along the surface of the wedge.
\newline
If block and wedge do not move relative to each other, differently from
the previous analysis, we have to consider two cases, according to
whether the block is on the brink of moving upward or downward along
the wedge's surface. We consider this latter situation, for which 
the balance of forces is $N (\sin\theta-\mu \cos\theta)=m a_{M}$ 
in the horizontal direction, and $N(\cos\theta+\mu \sin\theta)=mg$ in
the vertical direction. Thus the block does not slide ($a_{m} =0$) if
the wedge's acceleration is \cite{TT} 
\begin{equation}
a_{M}=g\frac{\sin\theta - \mu \cos\theta}{\sin\theta + \mu 
\cos\theta}.
\end{equation}
The corresponding relationship between accelerations is again a linear
one, as shown in {\it Fig. 3}, from which we deduce that
\begin{equation}
a_{m} =g(\sin{\theta} - \mu \cos{\theta})-
a_{M}(\cos{\theta} + \mu \sin{\theta}) \label{PRIM}.
\end{equation}
Note that Eq. (\ref{ACE}) holds when friction is present as well. So
from (\ref{ACE}) and (\ref{PRIM}) we obtain (See \cite{H} and numerous
examples on p. 86, 87 for a physical insight into the meaning of these
solutions)  
\begin{equation}
a_{M}=-\frac{mg\cos^{2}\theta (\tan\theta -\mu)}
{M + m - m \cos^{2}\theta (1+\mu\tan\theta)}
\end{equation}
and
\begin{equation}
a_{m}=\frac{(M+m)g\cos\theta(\tan\theta-\mu)}
{M + m - m \cos^{2}\theta (1+\mu\tan\theta)}.
\end{equation}
\noindent
We leave it to the reader to justify our considering the situation in
which the block is on the verge of sliding downward along the wedge
instead of upward.


\newpage
\begin{figure}
\centerline{
\epsfbox{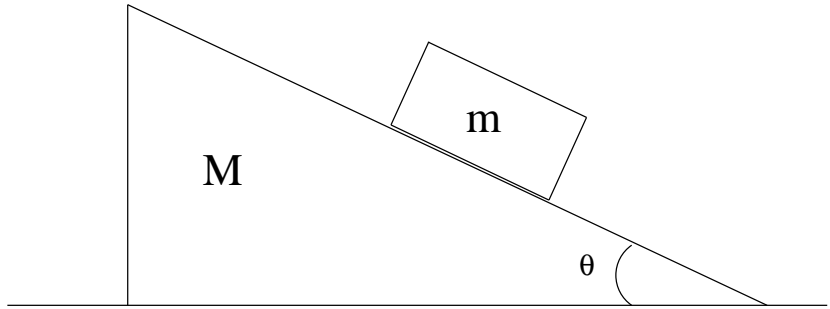}}
\caption{Block sliding on the wedge.}
\end{figure}

\begin{figure}
\centerline{
\epsfbox{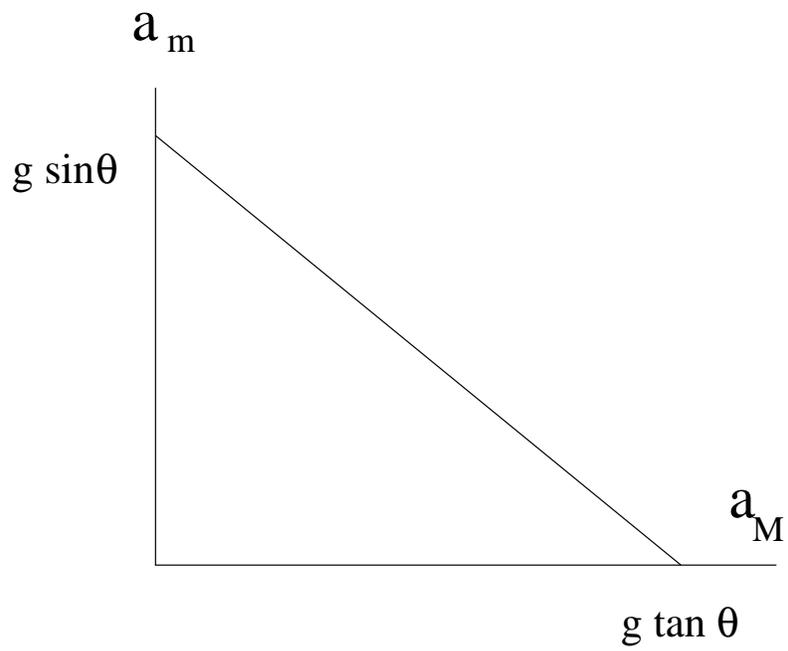}}
\caption{Relationship between $a_{m}$ and $a_{M}$.}
\end{figure}

\begin{figure}
\centerline{
\epsfbox{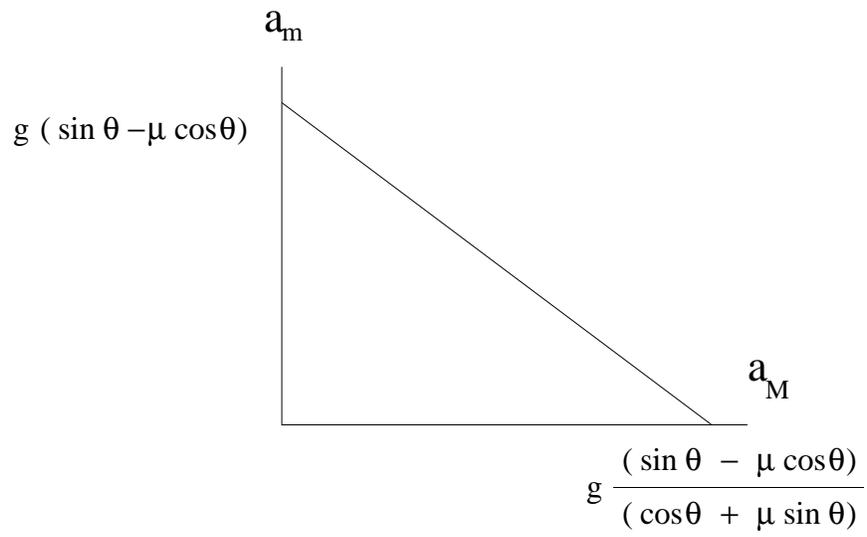}}
\caption{Relationship between $a_{m}$ and $a_{M}$ when $\mu \neq 0$.}
\end{figure}

\end{document}